\documentclass[12pt, table]{article}

\usepackage{graphicx}
\usepackage{authblk}
\usepackage[style=numeric,
            backend=biber,
            sorting=ynt
]{biblatex}
\usepackage{pdfpages} 
\usepackage{array}
\usepackage{longtable}
\usepackage{caption}
\usepackage{csquotes}
\usepackage{booktabs}
\usepackage{booktabs}
\usepackage{tikz}
\usetikzlibrary{shapes.geometric, arrows.meta, positioning}
\usepackage{graphicx}
\usepackage{float}
\usepackage{array}
\usepackage{enumitem}
\usepackage{tabularx}
\usepackage{fullpage}
\usepackage{tcolorbox} 
\usepackage{listings}
\usepackage{amsmath}
\usepackage{graphicx}
\usepackage{lipsum}
\usepackage{awesomebox}
\usepackage{smartdiagram}
    
\definecolor{arimlabs}{rgb}{0.0078, 0.0274, 0.4353}
\usepackage[
    colorlinks=true,
    linkcolor=arimlabs,        
    citecolor=arimlabs,        
    urlcolor=arimlabs       
]{hyperref}

\usetikzlibrary{positioning}
\hypersetup{colorlinks, citecolor=blue}
\usepackage{array}
\usepackage{booktabs}
\usetikzlibrary{positioning, arrows.meta}

\usepackage{listings}
\usepackage{lscape}
\definecolor{codegray}{rgb}{0.95,0.95,0.95}
\definecolor{codered}{rgb}{0.7,0.1,0.1}
\definecolor{codeblue}{rgb}{0.1,0.1,0.7}
\definecolor{codegreen}{rgb}{0,0.5,0.3}

\lstdefinestyle{custompython}{
    language=Python,
    backgroundcolor=\color{codegray},
    basicstyle=\ttfamily\footnotesize,
    keywordstyle=\color{codeblue}\bfseries,
    commentstyle=\color{codegreen}\itshape,
    stringstyle=\color{codered},
    showstringspaces=false,
    tabsize=4,
    breaklines=true,
    frame=single,
    rulecolor=\color{black},
    captionpos=b,
    numbers=none,
    escapeinside={(*@}{@*)}
}

\smartdiagramset{
    border color=none,
    uniform color list=teal!40 for 5 items,
    module x sep=3.75,
    back arrow distance=0.75
}
\title{
\hrule height 1mm
\vspace{25pt}
    Copyright in AI Pre-Training Data Filtering: Regulatory Landscape and Mitigation Strategies
\vspace{25pt}
\hrule height 1mm
}

\author[1,2]{Mariia Kyrychenko\textsuperscript{*†}}

\author[1,3]{Mykyta Mudryi\textsuperscript{*}}

\author[1,4]{Markiyan Chaklosh\textsuperscript{*}}
\affil[1]{ARIMLABS.AI}

\affil[2]{University of Texas at Tyler}
\affil[3]{Polish-Japanese Academy of Information Technology}
\affil[4]{University of the National Education Commission in Kraków}
\date{\today}

\affil[ ]{\textit {\{\href{mailto:mmudryi@arimlabs.ai}{mmudryi}, \href{mailto:mchaklosh@arimlabs.ai}{mchaklosh}\}@arimlabs.ai}}

\affil[ ]{\textit{\href{mailto:mariia.kyrychenko@outlook.com}{mariia.kyrychenko@outlook.com}}}

\usepackage[a4paper, top=0.5in, bottom=1in, left=1in, right=1in]{geometry}

\usepackage{fancyhdr}
\begin{document}

\begin{minipage}[h]{\textwidth}
    \maketitle
\begin{abstract}
     The rapid advancement of general-purpose AI models has increased concerns about copyright infringement in training data, yet current regulatory frameworks remain predominantly reactive rather than proactive. This paper examines the regulatory landscape of AI training data governance in major jurisdictions, including the European Union (or EU), the United States (or US), and the Asia-Pacific region. It also identifies critical gaps in enforcement mechanisms that threaten both creator rights and the sustainability of AI development. Through analysis of major cases we identified critical gaps in pre-training data filtering. Existing solutions such as transparency tools, perceptual hashing, and access control mechanisms address only specific aspects of the problem and cannot prevent initial copyright violations. We identify two fundamental challenges: pre-training license collection and content filtering, which faces the impossibility of comprehensive copyright management at scale, and verification mechanisms, which lack tools to confirm filtering prevented infringement. We propose a multilayered filtering pipeline that combines access control, content verification, machine learning classifiers, and continuous database cross-referencing to shift copyright protection from post-training detection to pre-training prevention. This approach offers a pathway toward protecting creator rights while enabling continued AI innovation.

\end{abstract}

\end{minipage}

\renewcommand{\thefootnote}{\fnsymbol{footnote}} 
\footnotetext[1]{These authors contributed equally.}
\footnotetext[2]{Correspondence should be addressed to: \href{mailto:research@arimlabs.ai}{research@arimlabs.ai}}
\setcounter{footnote}{1} 

\newpage

\section{Introduction}
The research community is concerned about the fact that many general-purpose AI (hereinafter GPAI) models are trained on data that could not be tracked, meaning that its provenance is unknown. These models
are allegedly trained on diverse compilations of text, image, and audio data scraped from the web, synthetically generated, or hand-curated \cite{longpre2024data}. \\

Data provenance is crucial for creators, developers, researchers, writers and others providing unique new resources. Without it, we face widespread copyright infringement, lack of creator consent, unreproducible research, and the exploitation of creative work without compensation or attribution issues that threaten both innovation and fair compensation in the digital economy.\\

This study employs comparative regulatory analysis across the EU, US, and Asia-Pacific jurisdictions, examining copyright frameworks, enforcement cases, and compliance mechanisms such as the General Data Protection Regulation (hereinafter GDPR) \cite{gdpr2016}, Copyright Directive \cite{eucopyright2019}, AI Act \cite{euaiact2024}, the Digital Services Act (hereinafter DSA) \cite{dsa2022} in EU, 
and case law and state-level legislation in the US. We analyzed lawsuit cases to identify enforcement gaps. We then evaluated existing technical solutions against open problems documented in academic databases such as Stanford's Technical AI Governance (hereinafter TAIG) Database \cite{stanford2024taig}.\\

Our analysis reveals two critical gaps in current copyright enforcement mechanisms for AI training data that are divided into technical and regulatory. First, \textbf{pre-training license collection and content filtering} emphasizes the impossibility of implementation of copyright clearance mechanisms. Second, the \textbf{verification mechanisms} contain regulatory and technical gaps. These challenges map directly to open questions in Stanford's Technical AI Governance Database \cite{stanford2024taig}: "How can licence collection be automated to prevent training on unlicensed data?", "How can problematic data be identified without full/direct access to the dataset?", and "How could the verification process for correct use of licensed data in AI model training be formalised?".\\

\section{Regulatory Landscape of AI Training Data}
To understand how the gaps mentioned above manifest in practice, we examine the regulatory frameworks that govern GPAI training data across major jurisdictions, beginning with the European Union's comprehensive approach to copyright and data protection.
\subsection{European Union}
In the EU, copyright infringement cases involving AI-generated content fall under the \textbf{EU Copyright Directive} \cite{eucopyright2019}. The \textbf{EU AI Act} \cite{euaiact2024} introduces transparency obligations for providers of general-purpose AI (GPAI) models. The \textbf{GDPR} \cite{gdpr2016} applies to privacy aspects, while the \textbf{DSA} \cite{dsa2022} to platform obligations.\\

\subsubsection{EU Code of Practice for GPAI Model Providers}
The EU Code of Practice for GPAI model providers serves as a voluntary framework designed to help GPAI model providers comply with the requirements of the EU AI Act. The Copyright Chapter, Measure 1.4 \cite[Copyright Chapter, Measure 1.4]{eu2024codeofpractice} suggests to mitigate the risk of copyright-infringing outputs. The way that AI Office can monitor infringement cases is through: \\
\textbf{1. Model Documentation\\}
GPAI model providers are required to describe the data filtering methodology used for GPAI in their Model Documentation. This method relies exclusively on trust rather than technical verification, and such verifications are carried out only when an infringement is reported or detected. \\
\textbf{2. Reporting Systems\\}
If the infringement case was reported, then the forensic investigation can be started by AI Office, so only then they can actually check if the training data filtering was enforced on GPAI or not.\\

Different information may be required from the GPAI model providers and it differs depending on the recipient, for example, AI Office, downstream provider, national competent authority.\\

Measure 1.3 of EU AI Code of Practice \cite[Copyright Chapter, Measure 1.3]{eu2024codeofpractice} states that the robots.txt protocol should delineate permissible and restricted directories for automated access, and Signatory non-compliance with these access restrictions constitutes a violation of established web crawling conventions and copyright reservation mechanisms. Beyond robots.txt other technical components and restrictions must be respected. This is not limited only to EU content, the restrictions must be respected everywhere. So if OpenAI (US) wanted to train its model on a pirated website (for example operated in Ukraine), it would not be able to sell its services to the EU content, as the providers should not train their model on any illegal content (Measure 1.2 of Copyright Chapter “Reproduce and extract only lawfully accessible copyright-protected content when crawling the World Wide Web”). The EU AI Act has extraterritorial reach, meaning that non-EU companies must comply with all EU AI Code of Practice requirements including copyright restrictions on training data from any global source, if they wish to access the EU market.\\
Signatories are also encouraged to take part in establishing communication with rightsholders (websites from which the data is scraped). Signatories must tell copyright holders how their web crawling system works and keep them informed about any changes.\\
Moreover, measure 1.3 states that if a GPAI model provider also provides a search engine (for example Google) then the two functions must be separated. For example, some website puts ‘no AI training’ in their metadata, but it still wants to be referenced when the user is looking for the information through a search engine. So, Google should not train its Gemini on the data from this website, but still show it to users through the search engine.\\

\subsubsection{EU Copyright Directive}
The Copyright Directive (EU) 2019/790 introduced exceptions for Text and Data Mining (TDM) which is the automated analysis of digital content to extract patterns, trends, and insights. Article 4 provides a general TDM exception that some interpret as potentially allowing AI companies to use copyrighted content for training purposes, though whether this exception applies to AI model training remains legally uncertain \cite{eparliamentstudy2025}. This provision includes a critical limitation: rightsholders can "opt out" by expressly reserving their rights through machine-readable means \cite[Article 4(3)]{eucopyright2019}. This creates a legal framework where TDM on copyrighted material is permitted by default, unless the copyright owner has technically implemented an opt-out mechanism on their website or platform. Under Article 4(3), rightsholders must use "machine-readable means, including metadata and terms and conditions of a website or a service" to reserve their rights against TDM. However, the directive provides limited guidance on what constitutes effective opt-out implementation, creating legal uncertainty for both AI developers and content creators. In Article 3 \cite[Article 3]{eucopyright2019} a separate, broader exception exists for research organizations and cultural heritage institutions conducting scientific research, which cannot be overridden by opt-out mechanisms. This creates a two-tiered system where commercial AI development faces potential restrictions that academic research does not.\\

\subsubsection{DSA}
The DSA mandates the creation of a first-of-its-kind public repository of content moderation decisions made by online platforms. This transparency mechanism could help track when platforms restrict AI scraping activities or remove content due to unauthorized use in AI training.\\
The DSA requires very large online platforms (VLOPs) to provide transparency about recommender systems and allow users to opt out of personalized recommendations. While not directly addressing AI training, this establishes precedent for algorithmic accountability that could extend to AI training data collection practices.\\

\subsubsection{GDPR}
The GDPR applies to the processing of personal data within its territorial scope, regardless of whether the data is obtained directly from data subjects or from publicly available or third-party sources. \cite{termly2024gdpr}. Recent enforcement actions, including proceedings involving OpenAI, have focused on inadequate justification of lawful bases, particularly failures in interest balancing, transparency, and purpose limitation in the context of AI training.\cite{dataprotectionreport2025edpb}.\\
Article 14 requires controllers to inform data subjects when processing personal data obtained from sources other than the data subject themselves, including publicly available data \cite{termly2024gdpr}. This creates obligations to respect contextual privacy expectations - for example, scraping should not occur where sites actively prohibit it via robots.txt or from platforms aimed at minors \cite{skadden2025cnil}.\\
Controllers must inform data subjects of their right to object (Article 21) when processing is based on legitimate interests, and in some cases provide specific notification and opportunity to object before processing occurs \cite{dataprotectionreport2024gdpr}.\\

A critical compliance gap emerges when new websites containing infringing content appear online but have not yet been identified or added to prohibited content databases. AI companies could unknowingly scrape this content for model training, raising questions about liability distribution between the AI company, content detection systems, and rights holders.\\
Under the EU AI Act's current framework, this scenario exposes multiple liability layers:\\
1. Regulatory Liability\\
The EU AI Code of Practice's "good faith" protection period ended in August 2025, meaning AI companies now face immediate regulatory enforcement for non-compliance. Companies can no longer rely on collaborative improvement periods with the AI Office and may face direct penalties for using prohibited training data.\\
2. Copyright Liability\\
Separate from regulatory enforcement, rights holders retain the ability to pursue copyright infringement claims in national courts under existing copyright law \cite{eucopyright2019}. The Code of Practice provides no shield against these civil lawsuits, creating parallel legal exposure.

\subsubsection{EU Enforcement Cases}
In March 2023, the Italian Data Protection Authority (Garante) launched an investigation into OpenAI's ChatGPT following concerns about its collection and processing of personal data. The investigation revealed that OpenAI had processed user data to train ChatGPT before the service's public release on November 30, 2022, without first identifying an adequate legal basis as required under GDPR \cite{LewisSlkin2025OpenAI}. \\

The Garante found multiple violations: OpenAI failed to notify Italian authorities of a data breach that occurred on March 20, 2023, which exposed chat histories and payment information of ChatGPT Plus subscribers during a nine-hour window and affected 440 Italian users. Additionally, OpenAI lacked sufficient age verification measures to protect minors and failed to fulfill transparency obligations toward users.\\

On March 30, 2023, the Garante issued an interim emergency decision ordering OpenAI to immediately stop processing personal data of Italian users through ChatGPT pending further investigation.  \cite{CliffordChance2023ChatGPT}.\\

Following discussions between OpenAI and the Garante in April 2023 \cite{Garante2023ChatGPTReinstated}, OpenAI proposed remedial measures which were analyzed and approved. On April 28, 2023, the Garante declared that OpenAI had implemented satisfactory measures including a privacy notice for users and non-users, and the right for all individuals in Europe to opt-out from data processing for algorithm training. On December 20, 2024, the Garante published Decision No. 755 (dated November 2, 2024) \cite{DataGuidance2024Garante}, imposing a €15 million fine on OpenAI for violations of GDPR Articles 5(1)(a), 5(2), 6, 12, 13, 24, 25, 32, and 33. Beyond the fine, the Garante required OpenAI to conduct a six-month public information campaign via radio, television, print media, and the Internet to inform users and non-users about how ChatGPT works, how data is collected, and how individuals can exercise their rights to object, erasure, and rectification under GDPR.

\subsection{The United States of America}
The US lacks the EU's comprehensive regulatory framework. While the EU Copyright Directive includes text and data mining (TDM) exceptions, their applicability to large-scale AI model training remains legally contested. In contrast, U.S. defendants rely on the open-ended fair use doctrine, which, like TDM, may be invoked to justify training activities, but is not a purpose-built exception for AI. However, several states within the US are implementing AI-specific copyright transparency requirements. Several U.S. states have proposed AI-specific copyright transparency obligations. For example, California Assembly Bill 412 (2025) \cite{california_ab412_2025} would have required developers of generative AI systems to document copyrighted materials used in training and to provide disclosure mechanisms for rights holders. However, the bill was not adopted, reflecting ongoing political and legal resistance to imposing training-data transparency obligations at the state level. Other states are enacting AI legislation, but most focus on employment discrimination, pricing algorithms, and consumer protection rather than copyright training data restrictions.

\subsubsection{US Enforcement Cases}
In December 2023, The New York Times filed a landmark lawsuit against OpenAI and Microsoft in the U.S. District Court for the Southern District of New York, alleging direct copyright infringement \cite{CopyrightAlliance2025Lawsuits, NYTimesComplaint2023}. The complaint alleges that the defendants used millions of the newspaper's copyrighted articles without permission or compensation to train their large language models, including ChatGPT and Microsoft's Bing Chat \cite{TechPolicyPress2025ANI}. The training dataset was massive in scale, with models trained on trillions of words—equivalent to a Microsoft Word document over 3.7 billion pages long \cite{HarvardLawReview2024NYT}.\\

The Times argues that this unauthorized use violates copyright law and undermines its business model by enabling users to bypass paywalls and obtain content substantially similar to Times articles without visiting the newspaper's website \cite{Stempel2023NYTimes}. The lawsuit demonstrates instances where AI models could reproduce content closely resembling specific Times articles, suggesting that training data was being memorized rather than truly transformed \cite{USC2025AICopyright}.\\

OpenAI and Microsoft defend their practices by invoking the fair use doctrine, a uniquely American legal principle that permits limited use of copyrighted material without permission under certain circumstances \cite{USC2025AICopyright}. In their answer to the amended complaint, OpenAI's attorneys argue that training AI models constitutes "paradigmatic transformative fair use" because the models learn statistical patterns rather than copying content for redistribution \cite{TheRegister2024OpenAITraining}. They contend that showing copyrighted books to AI models to teach them intelligence and language is analogous to human learning and does not constitute infringement because it extracts only word frequencies, syntactic patterns, and other statistical data rather than reproducing the works themselves \cite{TheRegister2024OpenAITraining}.\\

In November 2024, the court issued a significant ruling denying OpenAI's motion to compel evidence related to the New York Times' business practices or its employees' use of generative AI \cite{CopyrightAlliance2025Lawsuits}. The court determined that such discovery was irrelevant to the fair use analysis, rejecting OpenAI's attempt to argue that the Times' own use of AI undermined its claims. This ruling clarified that a fair use defense must stand on its own merits rather than on the plaintiff's behavior.\\

The case has been consolidated with other lawsuits filed by news organizations, including \textit{Daily News et al. v. Microsoft Corp. et al.} and \textit{The Center for Investigative Reporting v. OpenAI}, creating a unified proceeding that will address copyright issues affecting the entire news publishing industry \cite{CopyrightAlliance2025Lawsuits}.

\subsection{Comparison of EU and US approaches}
Unlike the EU's proactive regulatory enforcement through GDPR and the AI Act, the United States lacks comprehensive federal legislation governing AI training data \cite{USC2025AICopyright}. The US approach is characterized by:
\begin{enumerate}
\item \textbf{Case-by-case adjudication}: Rather than establishing clear rules, the US legal system relies on courts to determine whether AI training constitutes fair use through individual lawsuits \cite{CopyrightAlliance2025Lawsuits}.

\item \textbf{Industry self-regulation}: In the absence of federal mandates, AI companies have not been required to disclose training datasets or implement filtering systems, though some states are beginning to act \cite{USC2025AICopyright}.

\item \textbf{Emphasis on fair use balancing}: US courts will weigh four statutory factors: (1) purpose and character of use, (2) nature of the copyrighted work, (3) amount used, and (4) effect on the market for the original work \cite{USC2025AICopyright}.
\end{enumerate}

The New York Times case exemplifies the reactive, litigation-based US approach to AI governance, which contrasts sharply with the EU's proactive regulatory model. While the EU imposes immediate compliance obligations through GDPR and the AI Act—including transparency requirements, legal basis documentation, and substantial fines for violations—the US system allows AI companies to operate without such constraints while legal battles slowly proceed through the courts. The outcome of this case could fundamentally reshape how AI models are trained in the United States, potentially requiring licensing agreements with content creators or forcing companies to rely solely on public domain or licensed materials \cite{USC2025AICopyright}.

\subsection{Asian Countries}
\subsubsection{Japan}
Japan's revised Copyright Act (Article 30-4, effective January 1, 2019) allows broad rights to ingest and use copyrighted works for any type of information analysis, including AI training, for both commercial and non-commercial purposes \cite{privacyworld2024japan}. Unlike the EU, Japan allows commercial use and reportedly even permits training on "content obtained from illegal sites," with Japan's Education Minister indicating AI companies can use "whatever they want" for training \cite{privacyworld2024japan}.
\subsubsection{Singapore}
Singapore's Copyright Act Amendment of 2021 introduced Section 244, allowing use of copyrighted works for "computational data analysis" including commercial use, provided users have "lawful access" to the content \cite{techpolicy2024asia}.
\subsubsection{China}
China's "Interim Measures for the Management of Generative AI Services" (July 2023) requires that generative AI services "respect intellectual property rights," with Article 7.2 stipulating that AI providers must not infringe others' IP rights \cite{lexology2024apac,techpolicy2024asia}. China also requires transparency about training data use and has shown willingness to impose secondary liability on AI platforms for copyright infringement \cite{iic2025copyright,globallegal2025china}.
\subsubsection{South Korea}
Korea requires permission for using copyrighted works for AI training, as emphasized in Copyright Commission guidelines published in December 2023. Courts have ruled that web scraping infringes database rights under copyright law \cite{lexology2024apac,jtip2021copyright}.

\section{Challenges with AI Pre-Training Data Filtering}
While the regulatory frameworks discussed above establish important obligations, their practical implementation faces significant technical challenges that limit their effectiveness in preventing copyright infringement. The core challenge lies in the opacity of the training data filtering process. The main challenges identified in this paper are interconnected and collectively address the fundamental problem of pre-training data opacity. An analogy could be an e-commerce platform that scrapes product images from manufacturers' websites to populate its marketplace. Rights holders can readily identify their copyrighted images displayed on the platform and pursue infringement claims because the violation is directly visible in the final product. However, AI training presents a fundamentally different challenge: when copyrighted content is used to train language models, it becomes embedded within the model's weights through mathematical transformations. Unlike the e-commerce scenario where copied images remain visible, AI-generated outputs may reflect patterns learned from copyrighted material without reproducing exact copies, making it extremely difficult for rights holders to detect whether their specific content was used in training.\\

Therefore, filtering the data used to train GPAI is essential, but this process faces significant technical and regulatory challenges. 

\subsection{Pre-Training License Collection and Content Filtering}

Modern Large Language Models (LLMs) require training datasets containing billions of text tokens, images, and other media. The sheer volume of this data collection makes traditional copyright clearance mechanisms practically impossible to implement.\\

\textbf{Technical Gap:} The technical side of this problem lies in the difficulty of achieving comprehensive and accurate content filtering. Filtering systems can only exclude copyrighted content they can detect, yet copyright databases are incomplete, constantly updating, and fragmented across jurisdictions. Moreover, copyright protection extends to creative expression, not facts or ideas, requiring filtering systems to make nuanced legal determinations that are challenging even for human experts. Automated systems struggle to distinguish between copyrighted text and factual information, between protected creative works and legitimate quotations or fair use examples, and between original copyrighted material and publicly available reproductions. 

\textbf{Regulatory Gap:} the US Copyright Office noted, "it is not practically possible to obtain licenses for the volume and diversity of content necessary to power cutting-edge systems" \cite{uscopyrightoffice2025}.  New websites, blog posts, images, and other copyrighted materials appear online continuously, often before rights holders can implement or update restriction mechanisms. This temporal lag between content publication and rights assertion creates unavoidable exposure windows where AI companies may unknowingly incorporate restricted content into their training datasets.\\

\subsection{Verification Mechanisms Gaps}

The current enforcement paradigm relies heavily on post-training detection methods rather than pre-training prevention. Under current regulations like the EU AI Act, the AI Office primarily monitors compliance through complaint-driven investigations rather than proactive auditing of training datasets. \\

\textbf{Technical Gap: }The technical side of this problem lies in the absence of verification mechanisms for content filtering. While AI companies may implement systems to detect and exclude copyrighted material such as perceptual hashing or machine learning classifiers, which will be mentioned in the next section, there is no standardized way to audit whether these systems were correctly applied or successfully prevented copyright infringement. This creates scenarios where websites hosting unauthorized copyrighted content can monetize access to AI crawlers, who cannot distinguish between legitimate and pirated sources. The result is a verification paradox, for example, AI companies may pay for data access believing they have avoided copyright violations, while actually training on infringing content and compensating the wrong parties, with no mechanism for rights holders to detect or prove the violation.\\

\textbf{Regulatory Gap:}
On the regulatory side, verification mechanisms remain underdeveloped. The EU AI Code of Practice illustrates this limitation. While Measure 1.4 of the Copyright Chapter \cite[ Copyright Chapter, Measure 1.4]{eu2024codeofpractice}  requires GPAI providers to describe their data sources and data filtering methodologies in technical model documentation, it does not impose any obligation to implement verifiable controls or independent audits of those practices. Nor does it provide technical or procedural standards for assessing whether the documented filtering measures were actually applied or were effective in preventing the use of infringing content.

As a result, compliance is assessed largely on the basis of self-reported documentation. Verification typically occurs only in response to external triggers, such as regulatory complaints or litigation initiated by rights holders, rather than through systematic, proactive oversight. This reflects a predominantly reactive approach to pre-training data governance, in which deficiencies in filtering practices may remain undetected.\\

\begin{table}[h]
\renewcommand{\arraystretch}{1.5}
\centering
\begin{tabular}{|>{\raggedright\arraybackslash}p{3cm}|>{\raggedright\arraybackslash}p{3cm}|>{\raggedright\arraybackslash}p{4cm}|>{\raggedright\arraybackslash}p{4cm}|}
\hline
\textbf{Challenge Name} & \textbf{Core Problem} & \textbf{Technical Dimension} & \textbf{Regulatory Dimension} \\
\hline
Pre-Training Data Curation (3.1) & Scale makes comprehensive copyright management impossible & Filtering systems face accuracy limits: incomplete databases, inability to distinguish expression from facts, temporal detection lag & Licensing at scale is impractical; no legal framework exists for mass content licensing; temporal lag between publication and restriction \\
\hline
Verification Mechanisms (3.2) & Cannot confirm filtering prevented infringement & No standardized audit tools; cannot verify filtering implementation; cannot distinguish legitimate from pirated sources & Reactive complaint-driven enforcement; no proactive verification mandates; no third-party audit requirements \\
\hline
\end{tabular}
\caption{Copyright Challenges in AI Pre-Training Data}
\label{tab:copyright_challenges}
\end{table}

An example of mentioned above challenges is The New York Times' lawsuit against Microsoft and OpenAI \cite{nyt2023openai} which illustrates a serious category of potential copyright infringement. The Times alleges that AI models were trained on wholesale copying of Times articles, and that the resulting AI capabilities directly compete with the newspaper's core business model. The Times demonstrated instances where AI models could reproduce content substantially similar to specific Times articles, suggesting that training data patterns were being memorized rather than truly transformed. This raises significant problems as the training data used can't be tracked as well as after training it is impossible to access the dataset.\\

These three challenges correspond to the open research questions identified in Stanford's TAIG Database \cite{stanford2024taig} discussed before, demonstrating that the enforcement gaps we have identified reflect broader, unsolved problems in technical AI governance.

\section{Existing Mitigation Strategies}
To address these gaps in pre-training data filtering, several research initiatives and tools have emerged to help organizations better understand and manage their training datasets.\\

The \textbf{MIT Data Provenance Initiative} \cite{longpre2025data} represents a significant step forward in creating transparency around AI training data. This project conducted a systematic audit of more than 1,800 text datasets and released the \textbf{Data Provenance Explorer} \cite{dataprovenance2024explorer}, an open-source interactive tool that allows AI practitioners to trace the lineage of popular fine-tuning datasets and filter data based on specific license conditions. The tool enables users to generate human-readable data provenance cards for datasets, which significantly eases the manual task of curating and documenting extensive dataset compilations. The initiative found that many datasets lack proper documentation and attribution, with language coverage heavily skewed toward English and Western European languages, potentially introducing bias into AI models. This tool primarily serves three user groups: AI model builders who need to discover new datasets while filtering for licensing restrictions, dataset creators interested in tracking data origins for proper attribution, researchers and policymakers seeking to understand AI data transparency trends.\\

Complementing MIT's transparency-focused approach, Anthropic has developed breakthrough techniques for pretraining data filtering that go beyond basic compliance checking. Anthropic's methodology \cite{jindal2024anthropic} involves \textbf{multi-layered filtering systems} designed to identify and remove chemical, biological, radiological, and nuclear (CBRN) weapons-related information from pretraining data. Their approach targets dangerous dual-use knowledge that could enable malicious actors to develop weapons of mass destruction, rather than focusing on copyright violations. The filtering detects not only obvious CBRN content but also more subtle forms of potentially dangerous information, such as detailed synthesis procedures, weapons design specifications, and related technical knowledge. Their approach includes developing machine learning classifiers specifically trained to recognize harmful CBRN content patterns, with their best-performing "Prompted Constitutional classifier" achieving an F1 score of 0.96 in distinguishing harmful from harmless content. This proactive filtering represents a significant advancement over reactive post-hoc methods such as unlearning, safety fine-tuning, or output filtering, as it prevents dangerous content from entering the training pipeline rather than attempting to remove it after model deployment. While Anthropic's research addressed AI safety concerns related to CBRN weapons information, their methodology demonstrates the technical feasibility of sophisticated pre-training data filtering systems that can distinguish between target content and benign information with high precision. Their classifiers were trained specifically for safety applications, the underlying framework including their  classification approach, synthetic data generation techniques, and capability-preservation evaluation metrics could potentially be adapted to develop copyright specific filtering systems, though this would require substantial additional research and new training datasets focused on intellectual property rather than dangerous knowledge.\\

One more solution is the \textbf{"pay per crawl"} model \cite{cloudflare2024paypercrawl}, which allows content creators to have control over who accesses their work. If a creator wants to block all AI crawlers from their content, they should be able to do so. If a creator wants to allow some or all AI crawlers full access to their content for free, they should be able to do that, too. Moreover, pay per crawl grants domain owners full control over their monetization strategy. They can define a flat, per-request price across their entire site, meaning they get paid if an AI company wants to scrape the data from their website. Companies such as CloudFlare \cite{cloudflare2024paypercrawl}  have begun implementing and developing this approach, offering platforms for content creators to manage crawler access and monetize their data.\\

Another solution is \textbf{Named Entity Recognition} (hereinafter NER) \cite{li2022survey} which is used in AI Chatbots and security tools. It is a natural language processing technique that identifies and classifies specific entities in text into predefined categories such as: person names, organizations, locations, chemical compounds, works of art. NER could be adapted to identify copyrighted content by: detecting author names, publication titles, or publisher names; identifying trademarked product names or brand references; recognizing specific works (book titles, article headlines, song names); matching against databases of known copyrighted entities. However, NER alone is insufficient for copyright filtering because copyright protects expression, not just entity names.
For example, a text could discuss "The New York Times" without containing any Times content. Conversely, copyrighted text might not contain obvious entity markers, and, moreover, NER can't detect paraphrased or derivative content.\\

Also \textbf{perceptual hashing} \cite{sun2022deep,tang2024survey} is a great method that is already in use by many companies. Unlike cryptographic hashes (which change completely with any tiny modification), perceptual hashes create a "fingerprint" that remains similar even when content is slightly altered (resized, compressed, cropped, color-adjusted). The system can then compare these hashes to identify matching or near-matching content. While this technology is widely deployed by platforms like YouTube and Facebook \cite{titlow2016youtube} to identify copyrighted material after upload, it remains underutilized in the AI training data context. AI companies could theoretically check billions of scraped images and videos against copyright databases using perceptual hashing before including them in training datasets, but there's little evidence of systematic implementation, likely due to computational costs, incomplete database coverage, and legal ambiguity about whether using copyrighted images for training (without reproducing them in outputs) constitutes infringement.\\

\textbf{Have I Been Trained } \cite{HaveIBeenTrained} is a searchable database that allows artists and creators to check whether their images appear in major publicly documented AI training datasets like LAION-5B and LAION-400M, which were used to train models such as Stable Diffusion. Users can search by uploading images, entering artist names, or providing URLs to discover if their work has been included in training data, and can request opt-out from future dataset versions. However, the tool has significant limitations: it only covers datasets that companies have publicly disclosed (excluding proprietary datasets like OpenAI's), works retrospectively rather than preventing initial scraping, provides no legal enforcement mechanism, covers only images rather than text or code, and relies entirely on voluntary compliance from AI companies. This shows the reactive nature of current copyright protection mechanisms, creators can discover violations only after their work has already been scraped and used for training, and opt-out requests don't affect models already trained on the data, highlighting the gap between post-collection transparency tools and the proactive pre-training filtering mechanisms needed to prevent copyright infringement before it occurs.\\

\textbf{InnerProbe} \cite{ma2024innerprobe} presents a post-training framework for identifying copyright-related influences in Large Language Model outputs by leveraging internal multi-head attention (hereinafter MHA) mechanisms rather than relying solely on generated text analysis. Operating on already-trained models, the framework employs a two-component architecture: first, a lightweight Long Short-Term Memory (hereinafter LSTM) network processes MHA results in a supervised learning paradigm to perform sub-dataset contribution analysis, enabling precise attribution of generated outputs to specific copyrighted sources such as individual authors' works; second, a concatenated global projector trained via unsupervised contrastive learning facilitates the detection of non-copyrighted content by learning distinguishing patterns without requiring exhaustive labeled datasets. This methodology addresses critical limitations in existing approaches, which typically apply prompt engineering or semantic classification techniques that treat entire training corpora as uniformly copyrighted and struggle to trace specific source influences.Table below maps the identified challenges to existing solutions that partially address these problems.

\begin{table}[h]
\centering
\label{tab:challenges_solutions}
\begin{tabular}{|>{\raggedright\arraybackslash}p{5cm}|>{\raggedright\arraybackslash}p{8cm}|}
\hline
\textbf{Challenge} & \textbf{Existing Solutions} \\
\hline
\rule{0pt}{3ex}Pre-Training Data Curation (3.1) & 
\rule{0pt}{3ex}Pay-per-crawl systems
\newline
MIT Data Provenance Explorer
\newline
Perceptual hashing
\newline
Named Entity Recognition \\[1ex]
\hline
\rule{0pt}{3ex}Verification Mechanisms (3.2) & 
\rule{0pt}{3ex}InnerProbe
\newline
Have I Been Trained
\newline
MIT Data Provenance Explorer \\[1ex]
\hline
\end{tabular}
\caption{Mapping of Challenges to Existing Mitigation Strategies}
\end{table}

\section{Proposal}
A comprehensive solution to copyright infringement in AI pre-training data requires combining multiple complementary technologies into a unified filtering pipeline. We present a framework that would combine existing solutions, requires the enhancement of some of them and filters the pre-training data through different stages to prevent copyright violations before model training begins.\\

The system would begin with access control mechanisms like "pay-per-crawl", which must be enhanced with pre-crawl content analysis to verify ownership - preventing scenarios where pirated content aggregators monetize others' copyrighted materials. Before any data enters the training corpus, perceptual hashing systems would compare scraped images and videos against fingerprint databases, identifying copyrighted multimedia content even when slightly modified. For textual content, Named Entity Recognition would flag potential copyrighted sources by detecting author names, publication titles, and publisher identifiers, creating an initial filter that, while insufficient alone, reduces the corpus requiring deeper analysis. The filtered data would then pass through machine learning classifiers adapted from multi-method safety filtering frameworks, but retrained specifically to distinguish copyrighted from public domain text using synthetic training data and capability-preservation metrics. Throughout this process, transparency and data provenance tools would document data lineage and enable rights holders to audit training datasets, creating accountability mechanisms. Finally, all filtered data would be cross-referenced against copyright databases containing registered works, with continuous updates to capture newly published copyrighted materials. Together, these technologies create defense-in-depth, where each layer catches violations missed by others, transforming copyright protection from reactive post-training detection to proactive pre-training prevention. \\

\begin{figure}[H]
    \centering
    \includegraphics[width=0.8\textwidth]{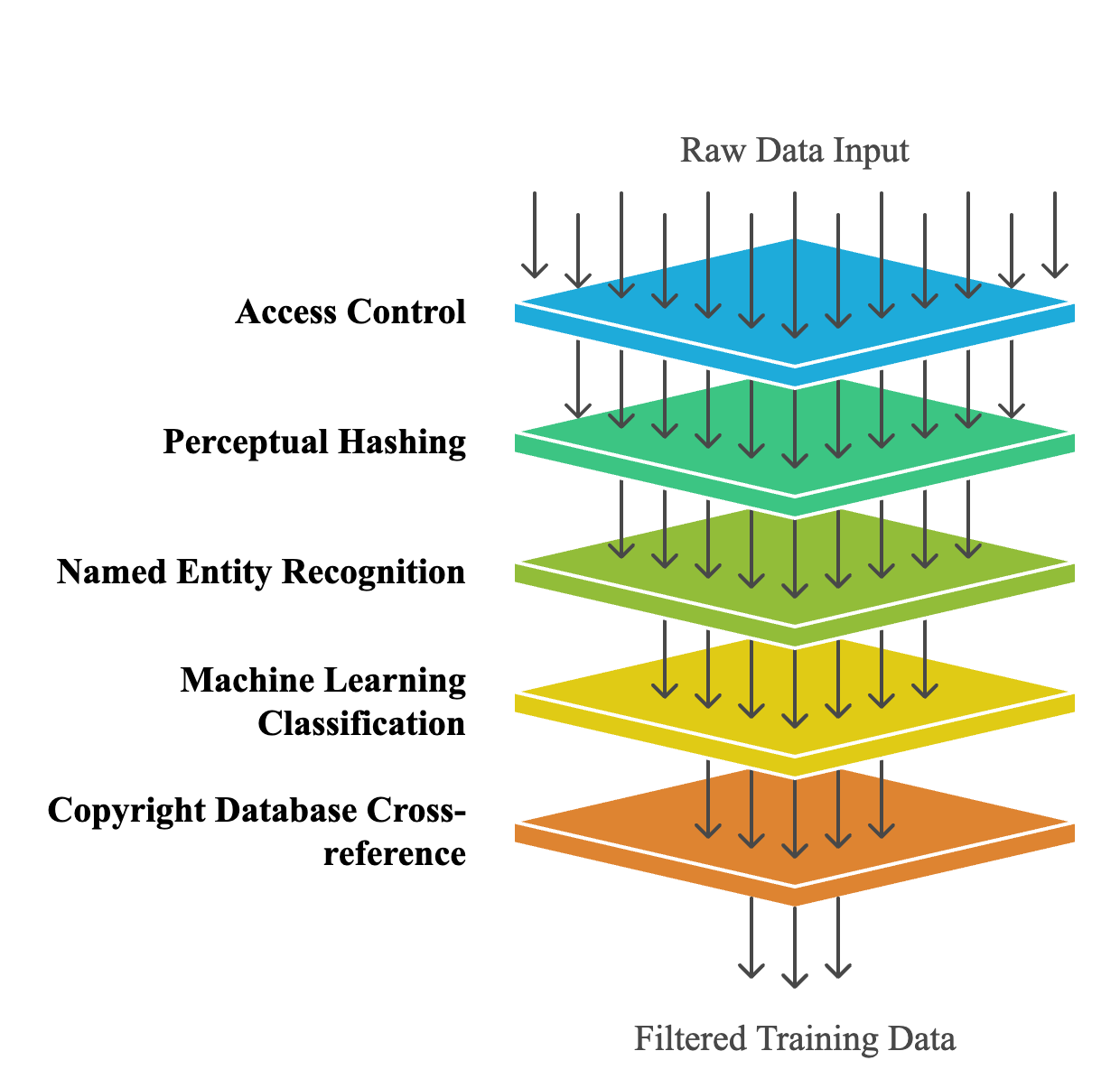}
    \caption{AI Copyright Infringement Prevention Funnel}
\end{figure}

While implementing such a system is feasible, it introduces meaningful technical and legal considerations. These include managing the computational scale of processing billions of documents across multiple filtering layers, maintaining continuously updated global copyright-reference databases, advancing detection techniques capable of identifying paraphrased or derivative content, and navigating an evolving international legal landscape regarding the use of copyrighted data for model training when outputs are not direct reproductions.\\

\section{Future Research}
Even with multi-stage filtering approaches, several important areas remain open for innovation and refinement, such as:
\begin{enumerate}
    \item Model Unlearning\\
    Models are already trained on vast quantities of unlicensed data collected before protective mechanisms existed. Future research should explore techniques for \textbf{dataset fingerprinting} to identify and trace the origins of training data, enabling retrospective auditing. \textbf{Watermarking techniques} embedded in original content could help detect unauthorized usage even after data has been incorporated into training sets. These watermarks could function as imperceptible patterns or signatures embedded directly into text, images, or other media that survive the training process and remain detectable in model outputs. For instance, statistical watermarking in text could involve subtle word choice patterns or syntactic structures that, when aggregated across multiple outputs, reveal the presence of specific copyrighted training material. Image watermarking could leverage frequency-domain modifications invisible to human perception but detectable through algorithmic analysis. The challenge lies in creating watermarks robust enough to survive the lossy compression and transformation inherent in neural network training, while remaining subtle enough to avoid degrading content quality or being easily removed by adversarial techniques.
    \item 
    Current solutions cannot effectively track data that circulates through third-party scrapers, academic institutions, or datasets traded outside controlled platforms. \textbf{Zero-knowledge proofs} offer an avenue for verifying data provenance without exposing the underlying content, allowing rights holders to confirm their content was used for training while preserving model confidentiality.
\end{enumerate}

\section{Conclusion}
This paper has examined the regulatory landscape governing AI training data across multiple jurisdictions and identified critical gaps in current enforcement mechanisms that threaten both creator rights and the sustainability of AI development. Regulatory frameworks increasingly require transparency and accountability, however, enforcement remains predominantly reactive rather than proactive. The EU's regulatory framework establishes clear obligations but rely heavily on complaint-driven investigations and post-deployment audits. The two core challenges identified are pre-training license collection and content filtering, where scale makes comprehensive copyright management impossible, and verification mechanisms, where we cannot confirm filtering prevented infringement. \\

Existing solutions demonstrate important progress but fundamental limitations. Transparency tools enable practitioners to trace dataset lineage and filter based on licensing restrictions, yet cannot prevent initial copyright violations. Pay-per-crawl approaches grant content creators monetization control, but cannot address already-collected data, indirect access through third parties, or the fundamental problem of detecting unauthorized content hosted on websites that have implemented the system. Searchable databases exemplify the reactive nature of current protections, creators discover violations only after their work has been scraped and potentially incorporated into deployed models, with opt-out requests affecting only future dataset versions. Even sophisticated post-training detection frameworks operate only after models have been trained and cannot prevent the initial infringement.\\

To address these limitations, we propose a multilayered filtering pipeline that shifts copyright protection from post-training detection to pre-training prevention. Our framework combines access control mechanisms enhanced with pre-crawl ownership verification, perceptual hashing systems for multimedia content identification, Named Entity Recognition for flagging potential copyrighted sources, machine learning classifiers retrained specifically to distinguish copyrighted from public domain content, and continuous cross-referencing against updated copyright databases. This defense-in-depth approach ensures each layer catches violations missed by others, transforming copyright enforcement from reactive to proactive. By implementing such filtering systems, supported by transparent data provenance documentation and enforced through clear regulatory mandates, we can achieve the dual objectives of protecting creator rights while enabling continued AI innovation.

\section*{\centering Acknowledgement}
The authors would like to thank \textbf{Hannah Khier} for his time and effort in reviewing this manuscript and for providing valuable comments and suggestions that helped improve its clarity and quality.
\noindent
\newline
\rule{\textwidth}{0.4pt}

\printbibliography

\end{document}